%% file: main.tex
\documentclass[letterpaper, 10 pt, conference]{ieeeconf}  

\IEEEoverridecommandlockouts                              

\usepackage[noadjust]{cite}
\usepackage{graphicx,epstopdf}
\usepackage[table]{xcolor}
\usepackage{amsmath,amsfonts,amssymb}
\usepackage{array}
\usepackage{longtable,tabularx}
\usepackage{multirow}
\usepackage{booktabs}

\newtheorem{theorem}{Theorem}

\newtheorem{lemma}{Lemma}
\newtheorem{definition}{Definition}
\newtheorem{remark}{Remark}

\newcommand{\U}{\mathcal{U}}
\newcommand{\T}{\mathcal{T}}
\newcommand{\G}{\mathcal{G}}
\renewcommand{\S}{\mathcal{S}}
\newcommand{\X}{\mathcal{X}}
\newcommand{\Y}{\mathcal{Y}}
\newcommand{\J}{\mathcal{J}}

\newcommand{\aK}[2]{\alpha_{#1}\left(\left\|#2\right\|\right)}

\newcommand{\R}[1]{\mathbb{R}^{#1}}
\renewcommand{\P}[1]{\mathcal{P}_{#1}}
\newcommand{\V}[1]{\mathcal{V}\left(#1\right)}
\newcommand{\Va}[1]{\tilde{\mathcal{V}}\left(#1\right)}
\DeclareMathOperator*{\argmin}{arg\,min}

\DeclareMathOperator{\sat}{sat}

\newcommand{\ra}[1]{\renewcommand\arraystretch{#1}}
\newcommand{\mysty}[1]{\cellcolor{blue!10}\textcolor{green!50!black}{\ifmmode\mathbf{#1}\else\textbf{#1}\fi}}

\usepackage{hyperref}
\hypersetup{
	colorlinks,
	linkcolor={red!80!black},
	citecolor={purple!80!black},
	urlcolor={blue!80!black}
}

\usepackage{tikz,tikz-3dplot,tikzinclude}
\usetikzlibrary{intersections,shapes,arrows,positioning}
\usetikzlibrary{shapes.geometric,calc,patterns}

\tikzset{BlockDiagram/.style={
	block/.style = {draw, rectangle, rounded corners, text centered, minimum height=1cm, minimum width=1cm, text width=2.5cm},
	sum/.style = {draw, circle, thick, minimum size=5mm, node distance=5mm, inner sep=0mm},
	input/.style = {coordinate},
	output/.style = {coordinate},
}} 

\title{\LARGE \bf
Nonlinear Control Allocation: A Learning Based Approach
}

\author{Hafiz Zeeshan Iqbal Khan$^{1}$, Surrayya Mobeen$^{2}$, Jahanzeb Rajput$^{1}$, and Jamshed Riaz$^{3}$
\thanks{$^{1}$Hafiz Zeeshan Iqbal Khan and Jahanzeb Rajput works at Centers of Excellence in Science and Applied
Technologies, Islamabad, Pakistan.}%
\thanks{$^{2}$Surrayya Mobeen is PhD student at Department of Aerospace, Toronto Metropolitan University, Toronto, ON, Canada.}%
\thanks{$^{3}$Jamshed Riaz is Professor at Department of Aeronautics \& Astronautics, Institute of Space Technology, Islamabad, Pakistan.}%
}

\begin{document}

\maketitle
\thispagestyle{empty}
\pagestyle{empty}

\begin{abstract}

Modern aircraft are designed with redundant control effectors to cater for fault tolerance and maneuverability requirements. This leads to aircraft being over-actuated and requires control allocation schemes to distribute the control commands among control effectors. Traditionally, optimization-based control allocation schemes are used; however, for nonlinear allocation problems, these methods require large computational resources. In this work, an artificial neural network (ANN) based nonlinear control allocation scheme is proposed. The proposed scheme is composed of learning the inverse of the control effectiveness map through ANN, and then implementing it as an allocator instead of solving an online optimization problem. Stability conditions are presented for closed-loop systems incorporating the allocator, and computational challenges are explored with piece-wise linear effectiveness functions and ANN-based allocators. To demonstrate the efficacy of the proposed scheme, it is compared with a standard quadratic programming-based method for control allocation.
\end{abstract}

\section{Introduction}
\input{Introduction.tex}

\section{Preliminaries}
\input{Preliminaries.tex}

\section{Problem Formulation}
\input{ProblemFormulation.tex}

\input{MainResults.tex}

\section{Aircraft Control Allocation: An Example}
\input{AircraftCA.tex}

\section{Conclusion} 
In this research work, a general nonlinear control allocation problem was posed from a different perspective, that is to seek a function that maps desired moments to control effectors. The allocation problem was converted to a machine learning problem and training methodologies for different types of allocation problems were presented, e.g., effectors prioritization, fault tolerance, reconfiguration, etc.

Two important results for the nonlinear control allocation problem were discussed in this work. Firstly, computational issues of performance parameters for piece-wise linear effectiveness functions with ANN-based allocators were discussed. Secondly, the conditions of closed-loop stability, with an allocator in the loop, in terms of maximum allocation error, were presented. There are a few future research avenues that need to be further explored regarding the development of efficient algorithms for detailed performance analysis of different allocators, e.g., volume computation of union of convex polytopes and the calculation of domain partitions for composition of piece-wise linear functions.

\addtolength{\textheight}{-12cm}   







\bibliography{References}
\end{document}

%% file: Introduction.tex
\begin{figure}
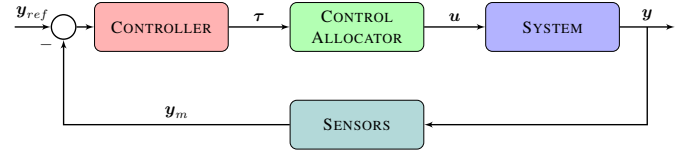

\centering \scalebox{0.62}{\includetikzgraphics[CAblkDiag]{TikzDatabase.tikz}}
\caption{Feedback-loop structure with \emph{control allocation}}
\label{fig:CABlockDiag}
\vspace{-13mm}
\end{figure}

Traditionally, aileron, elevator, and rudder are the three main effectors used in aircraft flight control, where each is intended for a particular degree of freedom. However, in newer designs, often many control effectors are used to provide better maneuverability \cite{Durham2017}, reconfiguration flexibility \cite{Khan2020_ECC} and required redundancy for fault-tolerance \cite{Dorsett1996}. The problem of distributing control commands to multiple control actuators is known as \emph{Control Allocation}, for which various methods have been proposed in literature \cite{Durham2017,Oppenheimer2011}. Advanced control techniques distinguish control command distribution from regulation. The baseline control law computes the desired control effort which is then distributed across the actuators by a specialized control allocator, as shown in Fig. \ref{fig:CABlockDiag}.

At present, a common practice is to use linear control allocation methods, such as direct allocation \cite{Durham1994,Bodson2002}, daisy chaining \cite{Buffington1996}, redistributed pseudo-inverse or cascaded generalized inverse \cite{Oppenheimer2011}, and methods based on linear programming \cite{Bodson2002,Buffington1999} and quadratic programming \cite{Bodson2002,Enns1998}. For most practical applications, optimization-based methods are preferred due to their accuracy and flexibility \cite{Johansen2013}. The control allocation problem of an aircraft is inherently nonlinear and coupled, especially the cross-channel effectiveness of surfaces can be highly nonlinear \cite{Niestroy2017,Rajput2018}. However, the problem of nonlinear control allocation is challenging due to the possibility of local minima and computational issues \cite{Johansen2013}. A conventional way is to treat the nonlinear mapping directly through sequential quadratic programming \cite{Poonamallee2004} or use locally affine approximations \cite{Rajput2023}, but the computational complexity could be enormous.

Artificial Neural Networks (ANNs) have proven to be excellent at approximating almost all nonlinear functions with arbitrary accuracy \cite{Csaji2001}. Initially, Grogan et al. \cite{Grogan1994} investigated an ANN-based linear control allocation and compared it with the direct allocation method for F/A-18 HARV aircraft. They concluded that the ANN technique was infeasible due to training constraints and accuracy problems. Quite recently, some research has been done on machine learning based control allocation schemes. Chen's use of RNNs for linear problems \cite{Chen2016}, Huan et al.'s application of deep auto-encoders for nonlinear problems \cite{Huan2018}, Vries et al.'s reinforcement learning for nonlinear allocation \cite{Vries2019}, Skulstad et al.'s constrained control allocator for dynamic ship positioning \cite{Skulstad2018}, and Kang et al.'s ANN-based dynamic allocation \cite{Kang2022} are examples of recent research on machine learning-based control allocation. Some recent studies also show significant improvements in the flight performance of multi-copters using ANN-based control allocation schemes \cite{Madruga2022,Ducard2023}. 
Most of the research that has already been done uses ANN-based approaches to solve linear control allocation problems. To the best of the authors' knowledge, there is limited information in the literature about an ANN-based framework to deal with challenges related to nonlinear control allocation.

In this research, we pose a general nonlinear control allocation problem in a different perspective, that is to seek a function that maps desired moments to control effectors.  Due to the excellent function approximation properties of ANN, we use them to approximate this mapping between desired moments and control effectors and convert the allocation problem to a machine learning problem. To demonstrate the efficacy of the proposed scheme, we compare its results with standard methods of control allocation in a high-fidelity closed-loop simulation of a miniature tailless flying-wing research aircraft \cite{Qu2017,Rajput2023}.

%% file: Preliminaries.tex
\subsection{Nonlinear Control Allocation}

The general nonlinear control allocation problem is defined as follows: find the control vector $u$ such that,
\begin{equation}\label{eq:01}
    \G(u,\sigma) = \tau_d
\end{equation}
where $u \in \U \subseteq \R{n}$ is actual input vector of the system being controlled, $\tau_d \in \R{m}$  is vector of desired moments (virtual control), $\sigma\in\S$ is some state vector, and $\G : \U\,\times\,\S \mapsto \T \subseteq \R{m}$ is control effectiveness mapping, and is assumed to be at least continuous in both $u$ and $\sigma$, where $\T$ (mentioned below) is known as Attainable Moment Set (AMS). Following are the some definitions we will be using in later sections:

\begin{definition}\label{def:AMS}
The \emph{pointwise} AMS is defined as,
\begin{equation}\label{eq:pwAMS}
  \T_\sigma \triangleq \G\left(\U,\sigma\right),\quad\forall\,\sigma\in\S
\end{equation}
and total or complete AMS, or just AMS, is defined as
\begin{equation}\label{eq:AMS}
  \T \triangleq \bigcup_{\sigma\in\S} \T_\sigma
\end{equation}
\end{definition}

The system of equations Eq. \eqref{eq:01} being underdetermined, usually possess multiple solutions, only if the constraints on input $u$ are strongly satisfied, otherwise no exact solution exists. Numerous methods have been proposed \cite{Oppenheimer2011,Johansen2013} to solve this constrained allocation problem, especially for cases when $\G$ is a constant linear map from $\U$ to $\T$, for example \emph{Redistributed Pseudo-Inverse (RPI)} or \emph{Cascaded Generalized Inverse (CGI)}, \emph{Direct-Allocation}, \emph{Daisy-Chaining}, and Optimization based methods.

For nonlinear $\G$, which is usually the case for modern over-actuated aircraft, it is a common practice to convert a nonlinear problem into a locally affine allocation problem at each sampling instant. At any sampling instant \eqref{eq:01} can be approximately written as;
\begin{equation}\label{eq:02}
  Gu \approx \bar{\tau}_d = \tau_d - \varepsilon
\end{equation}
where
\[
G = \left.\frac{\partial \G}{\partial u}\right|_{(u_0,\sigma_0)}, \text{and}\; \varepsilon = \G(u_0,\sigma_0) - G u_0
\]

Eq. \ref{eq:02} can be solved using any linear allocation method, depending upon the requirements, for which highly efficient tools are available. 

General nonlinear allocation problem \eqref{eq:01} can be posed as following weighted constrained optimization problem: given $\tau_d \in \T_\sigma$ and $\sigma \in \S$, solve
\begin{equation}\label{eq:03}
    \min_u \quad \big\|\G(u,\sigma) - \tau_d\big\|^2_W,\qquad
\end{equation}
subject to:
\[
    u \in \U
\]

where, $\|\cdot\|_W$ represents the weighted 2-norm and is defined as, $\|x\|_W^2 \triangleq x^\top W x$, where $W$ is a symmetric positive definite matrix. It must be noted that in this case, for a given $\tau_d$, optimization gives a vector $u\in\U$.

\subsection{Artificial Neural Networks (ANNs)} 
An Artificial Neural Network (ANN) consists of multiple layers of artificial neurons. Each neuron receives a vector from the previous layer as its input; and applies an affine transformation followed by a static nonlinear activation function. In this work the final layer doesn't have any activation function. Therefore, the network can be defined as,
\begin{equation}\label{eq:ANNPrem}
\begin{split}
   a_i &= \phi(\Theta_i a_{i-1} + b_{i}),\quad \forall\;i\in[1,N-1] \\
   a_N &= \Theta_N a_{N-1} + b_{N}
\end{split}
\end{equation}
If the network has $s_{i-1}$ neurons in $(i-1)$-th layer, and  $s_i$ neurons in $i$th layer, then  $\Theta_i\in\R{s_i\times s_{i-1}}$ is the weight matrix, and $b_i\in\R{s_i}$ is the bias term. $\phi:\R{s_i}\mapsto\R{s_i}$ is nonlinear activation function. The final output ($a_N$) of the nested functions showcase the successive connections of an $N$-layered ANN, where $a_0$ is the input vector.


%% file: ProblemFormulation.tex
The key idea of this research is, to try to find the mapping $u:\T\times\S\mapsto\U$, instead of single vector as in standard allocation problems. Though a similar work for linear case has already been done \cite{Grogan1994,Chen2016}, but in this work we have developed a generalized theory and posed it as a machine learning problem. Let's define a projection operator which will be used in subsequent development.

\begin{definition}[Projection Operator]
Consider a $k$-dimensional set $\mathcal{A}\subseteq\R{k}$, then the projection operator $\P{}:\R{k}\mapsto\mathcal{A}$ is defined as follows: 
\begin{equation}\label{eq:04}
    \P{\mathcal{A}}(x) \triangleq \argmin_{y\in\mathcal{A}} \|x - y\|^2
\end{equation}
\end{definition}

\begin{figure}
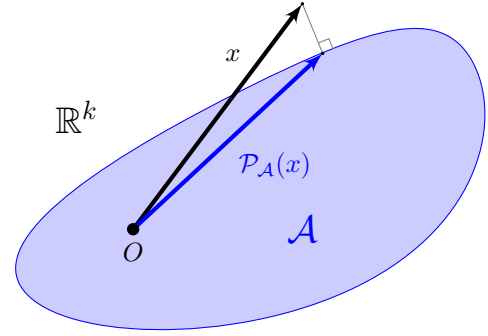

  \centering
  \scalebox{0.6}{\includetikzgraphics[ProjOpt]{TikzDatabase.tikz}}
  \caption{Projection Operator: An Illustration}
  \label{fig:ProjOpt}
  \vspace{-3mm}
\end{figure}

\begin{remark}
It must be noted that for a rectangular hyper-cubical set e.g. $\mathcal{A} = \{x\in\R{k} \mid x_{min} \leq x \leq x_{max} \}$, $\P{\mathcal{A}}$ reduces to vector saturation, i.e.
\[
\P{\mathcal{A}}(x) = \sat(x) \triangleq [\sat(x_1),\cdots,\sat(x_k)]^\top
\]
Also if $\mathcal{A} = \R{k}$ then $\P{\mathcal{A}}$ reduces to Identity map.
\end{remark}

\begin{lemma}\label{Lem:01}
Given an unconstrained optimization problem of the following form
\begin{equation}\label{eq:L01}
    x^{*}(y) = \argmin_{x\in\X} \quad f(x,y),\qquad \forall\,y \in \Y
\end{equation}
The optimal solution $x^{*}:\Y\mapsto\X$ can be equivalently considered as the solution of the following problem:
\begin{equation}\label{eq:L02}
    \min_{x(y)} \int\displaylimits_{\Y} f(x(y),y)\,\mathrm{d}y
\end{equation}
\end{lemma}
\begin{proof}
The necessary and sufficient conditions for \eqref{eq:L01} for any $y\in\Y$ can be written as
\[
\left.\frac{\partial f}{\partial x}\right|_{x=x^*} = 0,\quad \left.\frac{\partial^2 f}{\partial x^2}\right|_{x=x^*} > 0
\]
Similarly taking first and second variation of \eqref{eq:L02} and applying fundamental theorem of calculus of variations, yields same necessary and sufficient conditions, therefore, both \eqref{eq:L01} and \eqref{eq:L02} are equivalent.
\end{proof}

Using projection operator, we can re-write constrained optimization problem \eqref{eq:03} as following unconstrained one: given $\tau_d \in \T_\sigma$ and $\sigma \in \S$ solve
\begin{equation}\label{eq:NCAprob}
    \min_u \quad \big\|\G(\P{\U}(u),\sigma) - \tau_d\big\|^2_W
\end{equation}

Now applying Lemma 1 \cite{arXiv_preprint},  Eq. \eqref{eq:NCAprob} can be re-written as minimization of following functional
\begin{equation}\label{eq:COVprob}
    \J[u] = \int\displaylimits_{\T\times\S} \mathcal{L}(\tau,\sigma,u)\,\mathrm{d}\tau\mathrm{d}\sigma
\end{equation}
where
\[
\mathcal{L}(\tau,\sigma,u) = \Big\| \G \left(\P{\U}\left(u(\tau,\sigma)\right),\sigma\right) - \tau \Big\|^2_W
\]

Since generally $m > n$, so for most of the cases there doesn't exist any perfect $u(\tau,\sigma)$ for which $\J[u]=0$ . Our main goal is to find a map $u$, which minimizes $\J$.

Till now the development has been pretty general, and any numerical method for functional optimization problems can be used to solve \eqref{eq:COVprob}. However, if we discretize it over the domain, and consider ANN as a candidate for the map $u$, then we can equivalently pose \eqref{eq:COVprob} as the following learning problem, i.e. learn the network $\hat{u}$, while minimizing following cost over the network parameters.
\begin{equation}\label{eq:MLProb}
    \hat{\J} = \sum_{\T\times\S} \Big\| \G \left(\P{\U}\left(\hat{u}(\tau,\sigma)\right),\sigma\right) - \tau \Big\|^2_W
\end{equation}
It must be noted that the training of this network is not a standard \emph{supervised} machine learning problem.  Therefore, it requires further consideration as discussed subsequently.

\section{ANN-based Constrained Control Allocation}
\begin{figure}
  \centering
  \includegraphics[width=0.75\linewidth]{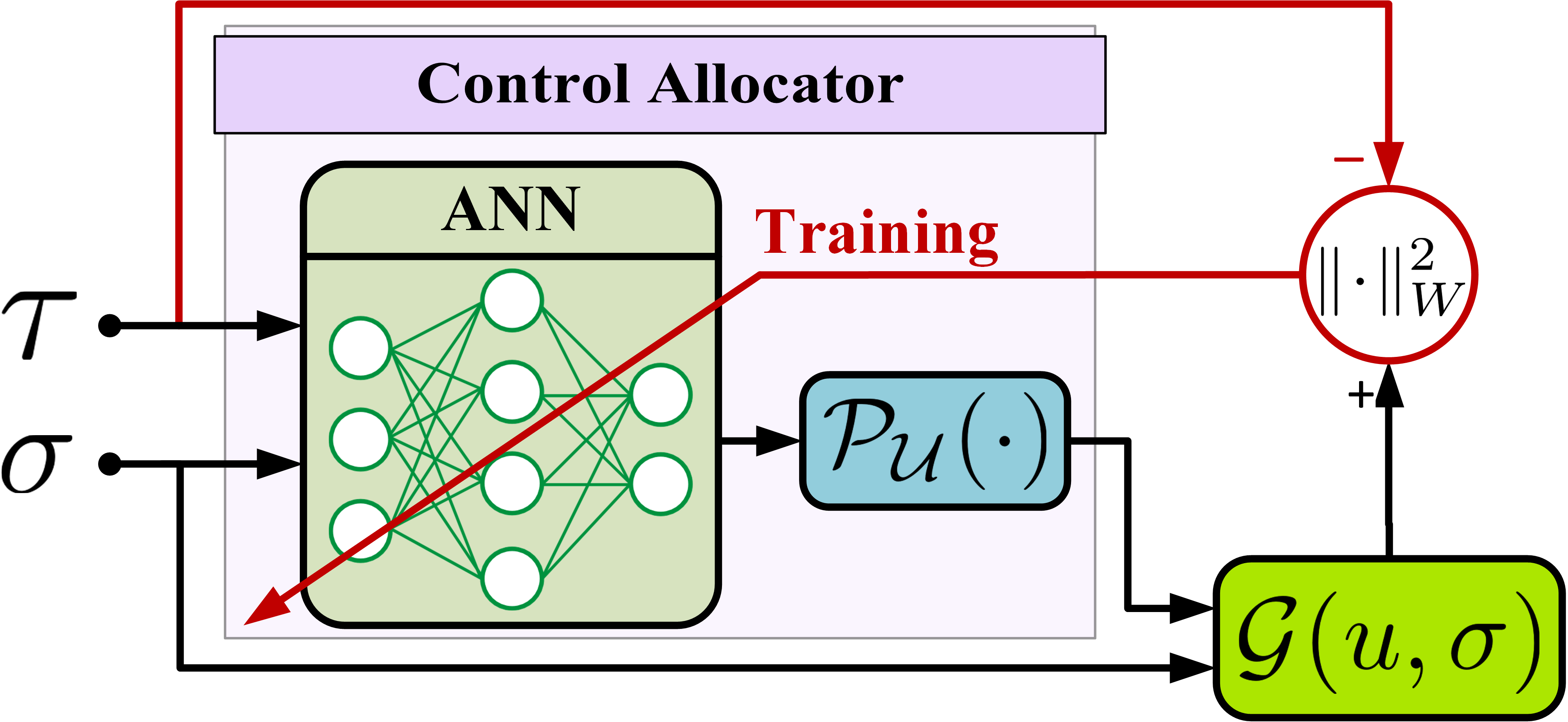}
  \caption{ANN Training Architecture}\label{fig:TrainingBL}
  \vspace{-3mm}
\end{figure}

To train a network, the first step is to obtain training data. For our case, given the sets $\U$ and $\S$ and map $\G:\U\times\S\mapsto\T$, we can generate random data-points in $\U\times\S$, and applying $\G$ gives corresponding points in $\T$. These random data-points can be generated using methods such as \emph{Latin Hypercube Sampling}. From this data neglect $\U$, and select data-points only in $\T\times\S$. This will be used as input of network. Also, no other output dataset is required as network is being trained according to the schematic shown in Fig. \ref{fig:TrainingBL}.

%% file: MainResults.tex
\subsection{Performance of Control Allocator}\label{sec:MainResults}
There are multiple criteria to compare the performance of control allocation methods. Most common are the \emph{Allocation Error} and \emph{Volume Ratio} of AMS to the actual AMS ($\T$) \cite{Durham2017}. In literature, allocation error is usually compared along a desired trajectory in $\T\times\S$ \cite{Durham1994}. In this section, we present a slightly different yet more general definition of \emph{Maximum Allocation Error (MAE)}. It can not only be used to compare different allocation methods but also for stability \& robustness analysis. Moreover, we define \emph{Volume Ratio} for general nonlinear control allocation problem. 
Efficient algorithms are available to compute \emph{Volume Ratio} for linear methods \cite{Durham2017}, but not for general cases.

Given the effectiveness map $\G:\U\times\S\mapsto\T$, sets $\U$, $\S$, and $\T$, and a control allocator $\hat{u}:\T\times\S\mapsto\U$, we define following performance measures for this allocator. Here, it must be noted that for following definitions it is not necessary for the control allocator $\hat{u}$ to be a function; it could just be an algorithm. However, it needs to be a deterministic one.

\begin{definition}
The \emph{Maximum Allocation Error (MAE)}  of an allocator $\hat{u}$ is defined as
\begin{equation}\label{eq:MAE}
  \Delta_{\hat{u}} \triangleq \max_{\sigma\in\S} \max_{\tau\in\T_\sigma} \Big\| \G\left(\P{\U}\left(\hat{u}(\tau,\sigma)\right),\sigma\right) - \tau \Big\|_1
\end{equation}
where $\|\cdot\|_1$ represents 1-norm.
\end{definition}

\begin{definition}
The $pointwise$ AMS of an allocator $\hat{u}$  can be defined as
\begin{equation}\label{eq:pwAMSM}
  \T^{\hat{u}}_\sigma \triangleq \G\left(\P{\U}\left(\hat{u}(\T_\sigma,\sigma)\right),\sigma\right),\quad\forall\sigma\in\S
\end{equation}
and similar to Definition \ref{def:AMS}, the total or complete AMS of an allocator $\hat{u}$  is defined as, $\T^{\hat{u}} \triangleq \bigcup_{\sigma\in\S} \T^{\hat{u}}_\sigma$.
\end{definition}

\begin{definition}
The \emph{Volume Ratio} of allocator $\hat{u}$ is defined as the ratio of volume of AMS of allocator and volume of actual AMS. This can be written as
\begin{equation}\label{eq:VR}
  \mathcal{R}^{\hat{u}} \triangleq  \min_{\sigma\in\S} \dfrac{\V{\T^{\hat{u}}_\sigma}}{\V{\T_\sigma}} 
\end{equation}
where $\mathcal{V}:2^{\R{m}}\mapsto\R{}$ represents the volume function.
\end{definition}

Now, let's consider a practical aircraft control allocation problem, where $\G$ is in general piecewise linear function.  Using this property in our ANN based allocation scheme yields following results:
\begin{theorem}
For piecewise linear $\G$, and ANN based allocator with only piecewise-linear activations (e.g. ReLU), and hyper-cubical input set $\U = \{u\in\R{m}| u_{min} \leq u \leq u_{max} \}$, the following holds:
\begin{enumerate}
\item The cost of \eqref{eq:MAE} is also a piecewise linear function.
\item The solution of \eqref{eq:MAE} can only be at the boundary of polytopes which divides the whole domain into finite regions and cannot be in interior of any of these polytopes.
\item If there are $\mathcal{N}$ polytopic regions for $\G\left(\U,\sigma\right)$ or $\G\left(\P{\U}\left(\hat{u}(\T_\sigma,\sigma)\right),\sigma\right)$, $\forall\,\sigma\in\S$ (a fixed $\sigma\in\S$), and $\T_i$ being the AMS (point-wise AMS) of $i$th region, then $\T_i$ is a convex polytope for all $i\in[1,\mathcal{N}]$, complete AMS (complete point-wise AMS) can be written as
\begin{equation}\label{eq:THMams_01}
      \T = \bigcup_{i\in[1,\mathcal{N}]} \T_i
\end{equation}
\end{enumerate}
\end{theorem}
\begin{proof} 
To prove the first statement, with hyper-cubical input set $\U$, since $\P{\U}(u) = \sat(u) \triangleq \min(\max(u,u_{min}),u_{max})$, therefore, $\P{\U}$ is piecewise linear. Each layer of an ANN can be written as $a_i = \phi(\Theta_i a_{i-1}+b_i)$ and since activation $\phi$ is assumed to be piecewise linear, so the complete ANN is piecewise linear. Moreover, the cost is 1-norm, which is defined as $\|x\|_1 = |x_1|+|x_2|+\cdots+|x_k|$, it is also, piecewise linear. Now, recalling that any combination of piecewise linear functions is also a piecewise linear function completes the proof of the first statement. \\
\noindent
For second and third statement, consider a single region (polytope) of the domain over which the cost is an affine function of inputs. The optimum (either maximum or minimum) of the function over the region cannot be in its interior, thus it must be on the boundary. Each region being a convex set when operated by an affine function results in a convex polytope (point-wise AMS). Therefore, complete AMS would be union of all these point-wise AMS of each region.
\end{proof}

Given $\mathcal{N}$ convex polytopes $\T_i, i\in[1,\mathcal{N}]$, the volume of their union can be computed by \emph{inculsion-exculsion} formula, but its computational cost is of order of $\mathcal{O}(n!)$. Another approach is to use the following identity:
\begin{equation}\label{eq:conv}
\mathcal{V}\left(\T\right) = \mathcal{V}\left(\mathcal{C}\right) - \mathcal{V}\left(\bigcap_{i\in[1,\mathcal{N}]}\mathcal{C} \setminus  \T_i\right)
\end{equation}
where $\T=\bigcup_{i\in[1,\mathcal{N}]} \T_i$, and $\mathcal{C}=\mathrm{Conv}\left(\T\right)$ represents convex hull. The second term in Eq. \eqref{eq:conv} is usually very small in magnitude as compared to first one, but requires a lot of computational resources. Therefore, we define an approximation of AMS volume as follows,
\begin{equation}\label{eq:approxVol}
\Va{\T} \triangleq \V{\mathrm{Conv}(\T)}
\end{equation}
and similarly, the approximate Volume Ratio of allocator $\hat{u}$ as,
\begin{equation}\label{eq:approxVR}
  \tilde{\mathcal{R}}^{\hat{u}} \triangleq \min_{\sigma\in\S} \dfrac{\Va{\T^{\hat{u}}_\sigma}}{\Va{\T_\sigma}}
\end{equation}
In the upcoming section of this paper, we have used global optimization techniques for computation of \emph{MAE} and Monte-Carlo based method for \emph{Volume Ratio} estimation.

\subsection{Closed Loop Stability Analysis}
Consider a system of the following form:

\begin{equation}\label{eq:sysOL}
  \dot{x} = f(x) + B(x)\overbrace{\G(u,\sigma(x))}^\tau
\end{equation}
where $x\in\X\in\R{k}$ is the state vector, $u \in \U \subseteq \R{n}$ is the true control input, $\tau\in\T\subseteq\R{m}$ is the virtual control input, $\G:\U\times\S\mapsto\T$ is control effectiveness, $B:\X\mapsto\R{k\times m}$ is a matrix with rank $m$, and $\sigma:\X\mapsto\S$ is a function of state; usually, it would be a subset of all states. Let's assume using some control design technique, we have designed a virtual control law $\tau = \P{\T_{\sigma(x)}}(k(x))$, which gives the \emph{ideal} closed-loop system
\begin{equation}\label{eq:isysCL}
  \dot{x} = f(x) + B(x)\P{\T_{\sigma(x)}}(k(x))
\end{equation}
After incorporating control allocator $\hat{u}$, we get the following \emph{actual} closed-loop system:
\begin{equation}\label{eq:asysCL}
  \dot{x} = f(x) + B(x)\G\left(\P{\U}\left(\hat{u}(k(x),\sigma)\right),\sigma\right)
\end{equation}

The basic idea of the following result is to treat the allocation error as non-vanishing but bounded perturbation at the input.

\begin{theorem}
Let the origin ($x=0$), be an asymptotically stable (AS) equilibrium of the ideal closed-loop system \eqref{eq:isysCL}, and let $V(x)$ be its Lyapunov function which satisfies
\begin{equation}\label{eq:THMsb_01}
\begin{split}
\aK{1}{x} \leq V(x) &\leq \aK{2}{x} \\
\left[\dfrac{\partial V}{\partial x}\right]^\top\left[f(x) + B(x)\P{\T_{\sigma(x)}}(k(x))\right] &\leq -\aK{3}{x} \\
\left\|\left[\dfrac{\partial V}{\partial x}\right]^\top B(x) \right\| &\leq \aK{4}{x}
\end{split}
\end{equation}
$\forall\,x \in \mathcal{D}$, where $\mathcal{D} = \{x\in\X \mid \|x\|\le r\}$, and $\alpha_1,\alpha_2,\alpha_3,\alpha_4$ are class $\mathcal{K}$ functions. Suppose that the MAE satisfies
\begin{equation}\label{eq:THMsb_02}
  \Delta_{\hat{u}} \le \theta \dfrac{\alpha_3\circ\alpha_2^{-1}\circ\alpha_1(r)}{\alpha_4(r)},\quad\forall\,x\in\mathcal{D},\,0<\theta<1
\end{equation}
Then, $\forall\,\|x(t_0)\| \le \alpha_2^{-1}\circ\alpha_1(r)$ the solution $x(t)$ of actual closed-loop \eqref{eq:asysCL} satisfies
\begin{equation}\label{eq:THMsb_03}
  \|x(t)\| \leq  \beta(\|x(t_0)\|,t-t_0), \quad\forall\,t_0\leq t \le t_0+T
\end{equation}
and
\begin{equation}\label{eq:THMsb_04}
  \|x(t)\| \leq  \rho(r), \quad\forall\,t \geq t_0+T
\end{equation}
for some class $\mathcal{KL}$ function $\beta$, and some finite $T$, where $\rho$ is class $\mathcal{K}$ function and is defined as
\begin{equation}\label{eq:THMsb_05}
  \rho(r) = \alpha_1^{-1}\circ\alpha_2\circ\alpha_3^{-1}\left(\frac{\Delta_{\hat{u}}}{\theta} \alpha_4(r)\right)
\end{equation}
\end{theorem}
\begin{proof}
For simplifying notations, in this proof we have used $\G = \G\left(\P{\U}\left(\hat{u}(k(x),\sigma)\right),\sigma\right)$, $\tau = \P{\T_{\sigma(x)}}(k(x))$, and $\|\cdot\|_p$ as $p$-norm. Using $V(x)$ as Lyapunov function for \emph{actual} closed-loop \eqref{eq:asysCL} we get,
\begin{equation*}
\begin{split}
\dot{V}(x) &= \left[\dfrac{\partial V}{\partial x}\right]^\top\left[f(x) + B(x)\tau + B(x)(\G-\tau)\right] \\
&\leq -\alpha_3(\|x\|) + \alpha_4(\|x\|)\|\G-\tau\|\\
&\leq -\alpha_3(\|x\|) + \alpha_4(\|x\|)\|\G-\tau\|_1,\;\; \because \|\cdot\|_2\leq\|\cdot\|_1\\
&\leq -\alpha_3(\|x\|) + \alpha_4(\|x\|)\Delta_{\hat{u}}\\
&\leq -(1-\theta)\alpha_3(\|x\|) -\theta \alpha_3(\|x\|) \\
& \qquad {}+ \alpha_4(\|x\|)\Delta_{\hat{u}}, \qquad\qquad 0<\theta<1\\
&\leq -(1-\theta)\alpha_3(\|x\|), \;\; \forall\;\|x\| \geq \alpha_3^{-1}\left(\frac{\Delta_{\hat{u}}}{\theta}\alpha_4(\|x\|)\right)\\
\end{split}
\end{equation*}
Applying Theorem 4.18 of \cite{Khalil2002} completes the proof.
\end{proof}

\begin{remark}
It should be noted that though the above results have been presented for a case of static controller $k(x)$, the same results can be applied to any continuous dynamic or observer based controller, even with multi-loop control architecture. First step in such a case would be to write the complete closed-loop system in the form of Eq. \eqref{eq:asysCL}, where $x$ represents all states (system, controller, observer - combined). Then this result can be applied directly.
\end{remark} 

\begin{remark}
In many practical scenarios, control allocation is used as much more than just distribution of control commands. It is also employed for reconfiguration in the occurrence of faults \cite{Khan2020_ECC}, and for prioritization of control effectors \cite{Buffington1996}. In case of control effectors' prioritization, daisy chaining approach is usually employed, which divides all effectors into multiple sets depending upon their priority. In our ANN-based approach this can be easily accomplished using multiple small ANNs for each set of effectors, and overall allocator can be obtained by appropriate stacking of these small networks.
\end{remark}

%% file: AircraftCA.tex
\ra{1.1}
\begin{table*}
\centering
\caption{Comparison of Different ANN Architectures}\label{tab:ANNArchs}
\begin{tabular}{lccccc}
\hline
\multicolumn{2}{c}{Control Allocator} & No. of Parameters & MSE & $R^2$  & MAE ($\Delta_{\hat{u}}$) \\ \cmidrule(rl){1-2}\cmidrule(rl){3-6}
\multirow{8}{*}{\begin{tabular}[c]{@{}c@{}}ANN\\Based\end{tabular}}
& 5.4.5            & 59          & $5.789\times10^{-3}$         & 0.9942         & 0.0325 \\
& 5.8.5            & 103         & $2.766\times10^{-3}$         & 0.9972         & 0.0300 \\
& \mysty{5.16.8.5} & \mysty{287} & $\mysty{1.115\times10^{-3}}$ & \mysty{0.9989} & \mysty{0.0227}\\
& 5.16.8.4.5       & 303         & $1.170\times10^{-3}$         & 0.9988         & 0.0226 \\
& 5.16.8.8.5       & 359         & $1.005\times10^{-3}$         & 0.9990         & 0.0215 \\
& 5.16.16.8.5      & 559         & $7.041\times10^{-4}$         & 0.9993         & 0.0214 \\
& 5.32.16.5        & 815         & $4.905\times10^{-4}$         & 0.9995         & 0.0188 \\
& 5.16.32.16.5     & 1263        & $4.471\times10^{-4}$         & 0.9996         & 0.0155 \\ \cmidrule(rl){1-2}\cmidrule(rl){3-6}
\multirow{2}{*}{\begin{tabular}[c]{@{}c@{}}QP\\Based\end{tabular}} 
& Polynomial & 684 & ${}-{}$ & ${}-{}$ & 0.0328 \\
& Spline     & 750 & ${}-{}$ & ${}-{}$ & 0.0776 \\ \hline
\end{tabular}
\end{table*}
\ra{0.7}

\subsection{Aircraft Specifications}\label{sec:V-A}
In this work, we have used an aerodynamic model of a small-scale tailless flying wing aircraft.  The comprehensive specifications of the aircraft can be found in \cite{Qu2017}.

The aircraft under discussion has six trailing edge surfaces ($\delta_1,\cdots,\delta_6$) and two (left/right) pairs of clam-shell surfaces ($\delta_{7U},\delta_{7L}$ and $\delta_{8U},\delta_{8L}$). Deflection of trailing edge surfaces in the downward direction is considered to be positive whereas in the upward direction, it is considered to be negative. In total, this aircraft has ten control surfaces. Any combination of these surfaces can be used to maneuver the aircraft. For this case, the authors have studied all trailing edge surfaces ($\delta_1,\cdots,\delta_6$) as a single elevator ($\delta_e$). This results in five independent control surfaces to be considered for the control allocation problem. The saturation limits for $\delta_e$ is $\pm$20 deg, for $\delta_{7U}$ and $\delta_{8U}$ is [0,40] deg, and for $\delta_{7L}$ and $\delta_{8L}$ is [-40,0] deg.


\subsection{Results \& Discussion}

Training data was randomly generated using \emph{Latin Hypercube} sampling. Multiple data sets comprising 0.05, 0.1, 0.2, and 0.4 million data points were generated. Each dataset was split with a ratio of 0.7, 0.15, and 0.15 into training, validation, and test data sets, respectively.  Keras API in Python was used to design and train the ANNs.  Adam optimizer was utilized as the optimization scheme, where the learning rate was initially kept at 0.005. The first and second moments were maintained at 0.9 and 0.999, respectively. Mean Squared Error (MSE) was kept as the loss function, while Root Mean Squared Error (RMSE) and R-squared ($R^2$) values were kept as metrics to measure the performance of the networks. Rectified Linear Unit (ReLU) was used as the activation function in all hidden layers due to its computational simplicity and effectiveness in handling of vanishing gradients. Every network was trained for 50 epochs with the data given to the network in multiple batches of 128 samples. Validation losses were also monitored simultaneously, and the learning rate was automatically reduced by a factor of 10\% if the validation loss did not improve after a \emph{patience} period of 5 epochs. Also, each network is trained 10 times and the best one is picked. It was also observed that the increase in the size of the data set, increases the accuracy, but after some threshold the benefit is minimal. Therefore, subsequently, results with only the largest data set (0.4 million points) will be discussed.



The first column in Table \ref{tab:ANNArchs} depicts different architectures of the ANNs studied. For example, `5.32.16.5' means that inputs and outputs remain five and have two hidden layers with 32 and 16 neurons, respectively. Out of five inputs for each network; the first three are desired moments and the last two are the angle of attack ($\alpha$) and sideslip angle ($\beta$). Outputs are deflection of five control surfaces ($\delta_e,\delta_{7U},\delta_{7L},\delta_{8U}$, and $\delta_{8L}$). Observing the values obtained of MSE and $R^2$ it can be noted that with an increase in hidden layers and number of neurons the network performance is improved.  It is also observed that the network performance of \textbf{`5.16.8.5'} is best suited in our opinion, as there is no substantial improvement in performance with increasing the number of hidden layers and the number of neurons after that.

\begin{figure}
  \centering
  \includegraphics[trim={8mm 1mm 1cm 2mm},clip,width=0.93\linewidth]{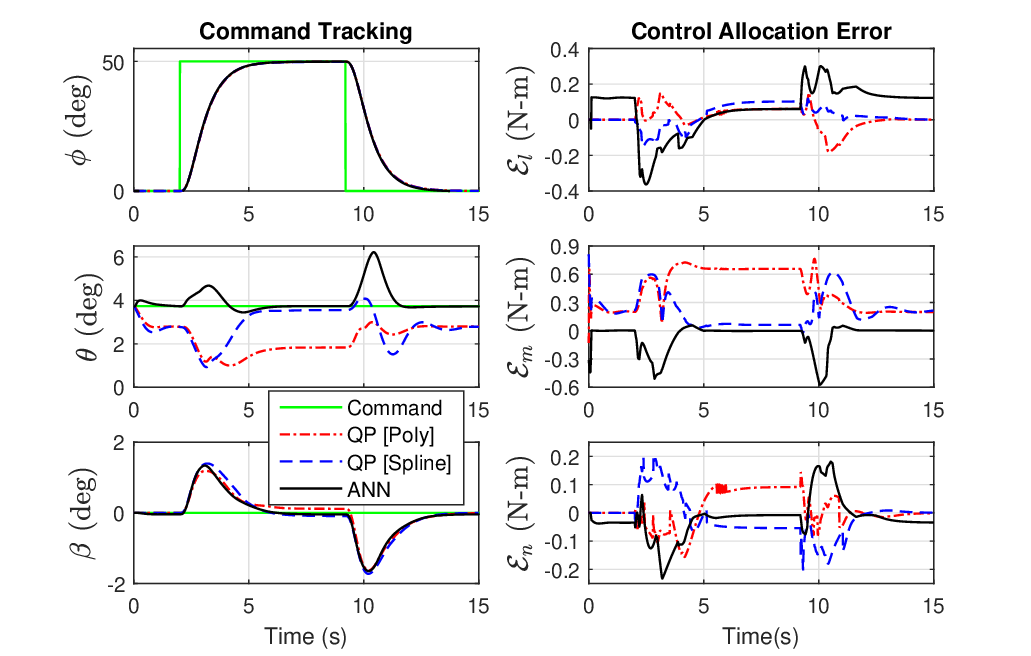}
  \caption{Comparison of Performance and Allocation Error}\label{fig:Comp}
  \vspace{-3mm}
\end{figure}

The proposed scheme's effectiveness is demonstrated through the implementation of the trained ANN in a 6-DoF high-fidelity nonlinear simulation of a flying wing aircraft. The ANN-based allocator is compared with an affine allocation method based on quadratic programming (QP) utilizing onboard models of control effectiveness based on Splines and Polynomials \cite{Rajput2023}. To assess the allocator's effectiveness, a U-turn maneuver in the simulation is generated, which demands significant control effort along all three axes. A 50-degree roll angle ($\phi$) command is applied, while the pitch angle ($\theta$) is maintained at its initial trim value (3.73 deg), and the sideslip angle ($\beta$) is kept zero. Figure \ref{fig:Comp} illustrates command tracking and control allocation errors for different schemes.

Despite similar allocation errors and closed-loop tracking performance between the ANN-based allocator and traditional methods, the proposed scheme is preferred due to its lower memory requirements and computational efficiency. The ANN-based method, relying on a few matrix operations, outperforms the QP-based method, which involves computing local slopes using onboard models and solving a quadratic program at each sampling instant. On average, the QP-based method took 1.608 msec with a Polynomial model and 9.013 msec with a Spline model, while the ANN-based method took only 0.02 msec, highlighting its significant computational efficiency. Note that these values are from MATLAB's \texttt{tic-toc} routine on a Core-i7 machine running at 1.7 GHz with MATLAB 2023a. During actual onboard implementation, values may differ but similar relative differences are likely to prevail.

